\documentclass[twocolumn,aps,prb]{revtex4}

\usepackage{amsmath,amssymb,amsfonts,bm}
\usepackage{graphicx,epstopdf}  
\usepackage{color}
\usepackage{lscape}
\usepackage{extarrows}
\usepackage{enumitem}
\usepackage{empheq}

\renewcommand{\d}{{\rm d}}

\newcommand{\M}{\mbox{\tiny M}}
\newcommand{\COP}{\mbox{\tiny COP}}
\newcommand{\POP}{\mbox{\tiny POP}}

\newcommand{\w}{\omega}
\newcommand{\wti}{\widetilde}

\newcommand{\B}{\mbox{\tiny B}}

\newcommand{\tS}{\mbox{\tiny S}}
\newcommand{\T}{\mbox{\tiny T}}

\newcommand{\dg}{\dagger}
\newcommand{\la}{\langle}
\newcommand{\ra}{\rangle}

\newcommand{\Sec}[1]{Sec.\,\ref{#1}}
\newcommand{\nl}{\nonumber \\}
\newcommand{\be}{\begin{equation}}
\newcommand{\ee}{\end{equation}}
\newcommand{\bsube}{\begin{subequations}}
\newcommand{\esube}{\end{subequations}}
\newcommand{\Eq}[1]{Eq.\,(\ref{#1})}
\newcommand{\Eqs}[1]{Eqs.\,(\ref{#1})}
\newcommand{\Fig}[1]{Fig.\,\ref{#1}}

\newcommand{\RN}[1]{%
  \textup{\uppercase\expandafter{\romannumeral#1}}%
}

\begin{document}

\title{Correlated driving--and--dissipation equation for non-Condon spectroscopy with the Herzberg--Teller vibronic coupling
}
\author{Jie Fang} \thanks{Authors of equal contributions}
\author{Zi-Hao Chen}
\thanks{Authors of equal contributions}
\author{Yao Wang}    \email{wy2010@ustc.edu.cn}
\author{Rui-Xue Xu}  \email{rxxu@ustc.edu.cn}
\author{YiJing Yan}
\email{yanyj@ustc.edu.cn}

\affiliation{Department of Chemical Physics, 
University of Science and Technology of China, Hefei, Anhui 230026, China}
\date{\today}
\begin{abstract}
Correlated driving--and--dissipation equation (CODDE) is an optimized complete second--order quantum dissipation approach, which is originally concerned with the reduced system dynamics only.
However, one can actually extract the hybridized bath dynamics from CODDE with the aid of dissipaton--equation--of--motion theory, a statistical quasi-particle quantum dissipation formalism.
Treated as an one--dissipaton theory, CODDE is successfully extended to deal with the Herzberg--Teller vibronic couplings in dipole--field interactions.
Demonstrations will be carried out on the non-Condon spectroscopies of a model dimer system.
\end{abstract}
\maketitle

\section{Introduction}\label{SecIntro}
Open quantum systems are ubiquitous in most fields of modern science.\cite{Wei21,Kle09,Bre02}
Quantum dissipation theory (QDT) dealing with the dynamics of open systems, such as quantum master equations,\cite{Nak58948,
Zwa601338,
Mor65423,
Red651,
 Lin76119,Che00729,Muk81509} is concerned with the reduced system dynamics. Its focus is the reduced density operator,  $\rho_{\tS}(t)\equiv {\rm tr}_{\B}\rho_{\T}(t)$, the bath subspace partial trace of the total composite density operator.
As a complete second--order quantum dissipation theory (CS-QDT),
the correlated driving--and--dissipation equation (CODDE)
method is of high efficiency.\cite{Xu029196,Xu037,Yan05187,Mo05084115} 
It is also found to be of high accuracy in its applicability range.\cite{ Xu046600, Xu13024106}
On the other hand, quasi-particle descriptions for baths
can provide a unified treatment on hybridized bath dynamics
and entangled system--bath properties.
This is right the point of the construction of the exact
dissipaton--equation--of--motion (DEOM) formalism
for Gaussian baths.\cite{Yan14054105,Zha15024112}
The DEOM adopts ``dissipatons'' as quasi-particles
associated with the bath coupling statistical dynamics.
For just the reduced system dynamics it recovers the hierarchical equations of motion (HEOM).\cite{Tan89101,Tan906676,Ish053131,Tan06082001,Yan04216,Xu05041103,Xu07031107,Din12224103,Tan20020901}
Beside, the underlying dissipaton algebra
enables DEOM an explicit theory for the hybridized bath dynamics.\cite{Yan14054105,Yan16110306,Wan20041102,Che21244105}

The CODDE considers explicitly
the driving--dissipation correlation,
which makes it convenient to plug--in the quasi-particle picture for
treating the hybridized bath dynamics. To observe that second--order QDTs are actually one-dissipaton theories, we have extended the CODDE to handle the Fano interference spectroscopies in previous work.\cite{Wan1894}
However, the Herzberg--Teller vibronic couplings in dipole-field interactions are still out of reach.
Theoretical study on the non-Condon vibronic coupling effect
needs to deal with the hybridization between the system
and the vibrational bath.
By the Herzberg--Teller approximation,
the bath degrees of freedom would be included in
the transition dipole moments.\cite{Lin743802,Zha16204109}
 In this work, we successfully extend CODDE to deal with the Herzberg--Teller vibronic couplings and simulate the corresponding non-Condon spectroscopies.

The rest of paper is organized as follows.
The DEOM theory is briefly introduced in
\Sec{sec2}.
In \Sec{sec3} we extend CODDE to deal with the Herzberg--Teller vibronic couplings with the aid of DEOM theory.
Simulations will be carried out on the non-Condon spectroscopies of a model dimer system in \Sec{sec4}.
We summarize this paper in \Sec{sec5}.

\section{The DEOM formalism}
\label{sec2}

\subsection{Prelude}
\label{thsec2A}

Let us start from the total composition Hamiltonian which assumes a form
\be\label{HT0}
  H_{\T}(t)=H_{\M}- \hat\mu_{\T}\epsilon(t) \equiv  H_{\tS} +h_{\B}
+\sum_{a}\hat Q^{\tS}_{a}\hat F^{\B}_{a}- \hat\mu_{\T}\epsilon(t),
\ee
where the external classical field $\epsilon(t)$ polarizes
both the system and bath environment.
For the Herzberg--Teller  coupling, we consider the total dipole moment to be
\be\label{muT0}
  \hat\mu_{\T}=\hat \mu_{\tS}\otimes\Big(1+\sum_{a}\hat \mu_{a}^{\B}\Big)=\hat \mu_{\tS}\otimes\Big(1+\sum_{a}\nu_{a}^{\B}\hat F_a^{\B}\Big).
\ee
Here, $H_{\tS}$ and $h_{\B}=\sum_j \w_j(p^2_j+x^2_j)$
are the Hamiltonians of an arbitrary quantum system and
a harmonic bath, respectively.
The system dissipative operators, $\{\hat Q^{\tS}_{a}\}$,
and dipole operators, $\hat \mu_{\tS}$,
are arbitrary and set to be dimensionless.
The bath hybridization operators, $\{\hat F^{\B}_{a}\}$,
are linear, i.e.\ $\hat F^{\B}_{a}=\sum_jc^a_jx_j$,
and the bath dipole components,
$\{\hat \mu^{\B}_a\}$, are
assumed to be $\hat\mu^{\B}_a\equiv \nu_{a}^{\B}\hat F_a^{\B}=\mu_a^{\B}\hat F^{\B}_{a}/(2\lambda_a)$
where $\lambda_a$ is the reorganization energy given later.
Throughout this paper we set $\hbar =1$ and
$\beta=1/(k_BT)$ with $k_{\B}$ being the Boltzmann constant
and $T$ the temperature.

The effects of such a Gaussian bath are characterized by
the bath coupling correlation functions,
\begin{align}
   C_{ab}(t)&\equiv \la \hat F^{\B}_a(t)\hat F^{\B}_b(0)\ra_{\B}
   \nl  &\equiv {\rm tr}_{\B}\big(e^{ih_{\B}t}\hat F^{\B}_a
                             e^{-ih_{\B}t}\hat F^{\B}_b
                             e^{-\beta h_{\B}}\big)
           /{\rm tr}_{\B}    e^{-\beta h_{\B}}.
\end{align}
They are related with
the bath coupling spectral densities,
\be\label{Jab_def}
 J_{ab}(\w) \equiv \frac{1}{2}\int^{\infty}_{-\infty}\!\! \d t\, e^{i\w t}
   \la [\hat F^{\B}_a(t),\hat F^{\B}_b(0)]\ra_{\B}=J^\ast_{ba}(\w)\,,
\ee
via the fluctuation--dissipation theorem as:\cite{Wei21,Yan05187}
\be\label{FDT}
C_{ab}(t)
=\frac{1}{\pi}\int^{\infty}_{-\infty}\!\!
  \d\w\, \frac{e^{-i\w t} J_{ab}(\w)}{1-e^{-\beta\w}}\,.
\ee
The associated bath reorganization energy is\cite{Wei21}
\be\label{reorg}
\lambda_{a}
=\frac{1}{2\pi}\int^{\infty}_{-\infty}\!\!
  \d\w\, \frac{J_{a}(\w)}{\w}\,.
\ee

To impose the quasi-particles on bath influences,
 exponential expansions of bath correlations are needed, i.e.\  
\be\label{FF_exp}
  C_{ab}(t)
    = \sum_{k=1}^{K} \eta_{abk} e^{-\gamma_{k} t}.
\ee
The coefficients in the exponents, $\{\gamma_k\}$,
required by \Eq{FDT}, should be either real or
complex conjugate paired.
Defining $\gamma_{\bar k}\equiv \gamma^{\ast}_k$,
there would be
\be\label{FF_exp_rev}
C_{ab}(-t)=C_{ba}^\ast(t)=\sum_{k=1}^{K} \eta^{\ast}_{abk} e^{-\gamma^{\ast}_{k} t}
= \sum_{k=1}^K \eta^{\ast}_{ab{\bar k}}e^{-\gamma_k t}.
\ee
The following quasi-particle descriptions on baths are just introduced
on basis of \Eq{FF_exp} with the property of \Eq{FF_exp_rev}.
Here, we assume $\{\gamma_k\}$ independent of $a,b$. Extension to general cases is straightforward.\cite{Xu07031107,Che21244105}

\subsection{The dissipaton algebra}
\label{thsec2B}

The DEOM construction starts from the dissipaton decomposition
on the hybridization bath operator,\cite{Yan14054105,Yan16110306}
\be\label{F_in_f}
 \hat F^{\B}_a = \sum_{k=1}^K \hat f_{ak},
\ee
where the involving statistically independent dissipatons satisfy
\be\label{ff_corr}
\begin{split}
 \la \hat f_{ak}(t)\hat f_{bj}(0)\ra_{\B}
 &=\delta_{kj}\eta_{abk}e^{-\gamma_k t},
\\
 \la \hat f_{bj}(0)\hat f_{ak}(t)\ra_{\B}
 &= \delta_{kj}\eta^{\ast}_{ab{\bar k}}e^{-\gamma_k t},
\end{split}
\ee
to reproduce \Eqs{FF_exp}
and (\ref{FF_exp_rev}).
It thus leads to
the generalized diffusion equation of dissipatons,\cite{Yan14054105,Yan16110306}
\be\label{diff}
 {\rm tr}_{\B}\Big[\Big(\frac{\partial}{\partial t} \hat f_{ak}\Big)_{\B}\rho_{\T}(t)\Big]
 =-\gamma_{k}\,
    {\rm tr}_{\B}\big[\hat f_{ak}\rho_{\T}(t)\big],
\ee
where
\be\label{hB_Heisenberg}
  \Big(\frac{\partial}{\partial t} \hat f_{ak}\Big)_{\B}=-i[\hat f_{ak},h_{\B}],
\ee
obeys the Heisenberg equation
of motion in bare bath.

The dissipaton density operators (DDOs) that constitute the DEOM are 
defined as:\cite{Yan14054105,Yan16110306}
\be\label{DDO}
 \rho^{(n)}_{\textbf{n}}(t)\equiv {\rm tr}_{\B}\Big[
  \Big(\prod_{ak} \hat f^{n_{ak}}_{ak}\Big)^\circ
  \rho_{\T}(t)\Big].
\ee
Here, $\textbf{n}=\{n_{ak}\}$, $n=\sum_{ak} n_{ak}$,
and $(\cdots)^{\circ}$ denotes the \emph{irreducible} product.
For bosonic dissipatons,
 $(\hat f_{ak}\hat f_{bj})^{\circ}=(\hat f_{bj}\hat f_{ak})^{\circ}$,
and $n_{ak}=0,1,\cdots$ can be viewed as
the occupation numbers of
individual dissipatons.
For the later construction of DEOM,
denote also $\rho^{(n\pm 1)}_{{\bf n}^{\pm}_{ak}}$ as the associated
$(n\pm 1)$-DDOs, with
${\bf n}^{\pm}_{ak}$ differing from ${\bf n}$ only
at the specified $\hat f_{ak}$-disspaton occupation number,
$n_{ak}$, by $\pm 1$.
The involving generalized Wick's theorem on 
dissipatons thus reads,\cite{Yan14054105,Yan16110306} cf.\ \Eq{ff_corr},
\begin{align}\label{Wick1}
 &\quad {\rm tr}_{\B}\Big[\Big(\prod_{ak} \hat f^{n_{ak}}_{ak}\Big)^\circ
   \hat f_{bj} \rho_{\T}(t)\Big]
\nl&=
 \rho^{(n+1)}_{{\bf n}^{+}_{bj}}(t)+\sum_{ak} n_{ak}\la\hat f_{ak}(0^+)\hat f_{bj}(0)\ra_{\B}
   \rho^{(n-1)}_{{\bf n}^{-}_{ak}}(t),
\end{align}
and
\begin{align}\label{Wick2}
 &\quad {\rm tr}_{\B}\Big[\Big(\prod_{ak} \hat f^{n_{ak}}_{ak}\Big)^\circ
   \rho_{\T}(t)\hat f_{bj} \Big]
\nl&=
 \rho^{(n+1)}_{{\bf n}^{+}_{bj}}(t)+\sum_{ak} n_{ak}\la\hat f_{bj}(0)\hat f_{ak}(0^+)\ra_{\B}
   \rho^{(n-1)}_{{\bf n}^{-}_{ak}}(t).
\end{align}
These dissipaton algebra will be used 
in the later DEOM construction in the presence of non-Condon external field excitation.

\subsection{The DEOM construciton}
\label{thsec2C}

 The DEOM can now be readily constructed
by applying $\dot{\rho}_{\T}(t)=-i[H_{\T}(t),\rho_{\T}(t)]$,
to the total composite density operator in \Eq{DDO};
i.e.,
\be\label{DDO_dot}
 \dot\rho^{(n)}_{\textbf{n}}(t)= -i\, {\rm tr}_{\B}\Big\{
  \Big(\prod_{ak} \hat f^{n_{ak}}_{ak}\Big)^\circ
  [H_{\T}(t),\rho_{\T}(t)]\Big\}.
\ee
By \Eq{diff} with \Eq{hB_Heisenberg} 
and \Eqs{Wick1}--(\ref{Wick2}),
the final DEOM formalism, with the total Hamiltonian given in \Eq{HT0},
is resulted as
\begin{align}\label{DEOM}
 \dot\rho^{(n)}_{\bf n}=
& -\Big[i{\cal L}(t)+\sum_{ak} n_{ak}\gamma_k\Big]\rho^{(n)}_{\bf n}
\nl&
  -i\sum_{ak}\big[{\cal A}_{a}-{\cal D}_{a}\epsilon(t)\big]\rho^{(n+1)}_{{\bf n}_{ak}^+}
\nl&
  -i\sum_{ak}n_{ak}\big[{\cal C}_{ak}-{\cal D}^\prime_{ak}\epsilon(t)\big]
   \rho^{(n-1)}_{{\bf n}_{ak}^-} .
\end{align}
Here,
\bsube\label{calAC}
\begin{align}\label{calL}
&{\cal L}(t)\hat O \equiv [H_{\tS}-\hat\mu^{\tS}\epsilon(t),\hat O], \\ \label{calA}
&{\cal A}_a\hat O\equiv [\hat Q^{\tS}_a,\hat O],\qquad
 {\cal D}_a\hat O\equiv \nu_{a}^{\B}[\hat\mu_{\tS},\hat O], \\ \label{calC}
&{\cal C}_{ak}\hat O
\equiv \sum_b \Big(\eta_{abk}\hat Q^{\tS}_a\hat O
        -\eta^{\ast}_{ab\bar k}\hat O\hat Q^{\tS}_a\Big), \\ \label{calD}
&{\cal D}^\prime_{ak}\hat O \equiv 
       \sum_b \nu_b^{\B}\big(\eta_{abk}\hat \mu_{\tS}\hat O
        -\eta^{\ast}_{ab\bar k}\hat O\hat \mu_{\tS} \big).
\end{align}
\esube
Therefore, the DEOM provides a unified approach to study
the system--bath entangled polarizations.

\subsection{Spectroscopies}
Let us first consider the non-Condon polarization,
\be\label{PTt_def}
  P_{\T}(t)={\rm Tr}[\hat\mu_{\T}\rho_{\T}(t)] =
    {\rm tr}_{\tS}{\rm tr}_{\B}\Big[\hat \mu_{\tS}\Big(1+\sum_{a}\hat \mu_{a}^{\B}\Big)\rho_{\T}(t)\Big].
\ee
Applying \Eqs{DDO} and (\ref{Wick1}),
the DEOM--evaluation on it reads \cite{Zha16204109,Che21244105}
\be\label{PTt_DDOs}
 P_{\T}(t)={\rm tr}_{\tS}\Big[\hat\mu_{\tS}\rho^{(0)}(t)
   +\sum_{ak}\nu^{\B}_a\hat\mu_{\tS}\rho^{(1)}_{ak}(t)\Big].
\ee
The dipole--dipole correlation function is\cite{Muk95}
\be\label{dipole_corr}
  \la \hat\mu_{\T}(t)\hat\mu_{\T}(0)\ra
\equiv
  {\rm Tr}(\hat\mu_{\T} e^{-i{\cal L}_{\M}t}\hat\mu_{\T}\rho^{\rm eq}_{\T}).
\ee
Here, ${\cal L}_{\M}\,\cdot\,\equiv [H_{\M},\cdot\,]$
and $\rho^{\rm eq}_{\T}=e^{-\beta H_{\M}}/{\rm Tr}e^{-\beta H_{\M}}$
are the total matter Liouvillian
and the thermal equilibrium density operator
in the absence of external field, respectively [cf.\,\Eq{HT0}].
The DEOM correspondence to $\rho^{\rm eq}_{\T}$
is just ${\bm\rho}_{\rm eq}\equiv \big\{\rho^{(n)}_{{\bf n};{\rm eq}}\big\}$,
the steady--state solutions to the field--free DEOM (\ref{DEOM}).
The DEOM evaluation on the dipole correlation
function, \Eq{dipole_corr}, is as follows.\cite{Wan20041102,Che21244105}

\noindent (i) Start with the aforementioned correspondence
of $\rho^{\rm eq}_{\T}\Rightarrow
\big\{\rho^{(n)}_{{\bf n};{\rm eq}}\big\}$,
by evaluating the steady--state solutions to the field--free \Eq{DEOM};

\noindent (ii) Identify $\hat\mu_{\T}\rho^{\rm eq}_{\T}\Rightarrow
\{\rho^{(n)}_{\bf n}(t=0; \hat \mu_{\T})\}$ by using \Eq{DDO},
and obtain  \cite{Zha15024112}
\begin{align}\label{rhon_t0}
 &\quad\, \rho^{(n)}_{\bf n}(t=0; \hat \mu_{\T})
\equiv {\rm tr}_{\B}\Big[
  \Big(\prod_{ak} \hat f^{n_{ak}}_{ak}\Big)^\circ
  \big(\hat\mu_{\T}\rho^{\rm eq}_{\T}\big)\Big]
\nl&
 \!\!=\!\hat \mu_{\tS}\Big[\rho^{(n)}_{{\bf n};{\rm eq}} \!+\!
  \sum_{ak} \nu^{\B}_a\rho^{(n+1)}_{{\bf n}^{+}_{ak};{\rm eq}}
  \!+\!\sum_{ak,b} \nu^{\B}_{b} n_{ak}\eta_{abk}
    \rho^{(n-1)}_{{\bf n}^{-}_{ak};{\rm eq}}\Big].\!
\end{align}
The second identity is obtained by using \Eq{muT0} for $\hat\mu_{\T}$,
followed by the generalized Wick's theorem, \Eq{Wick1};

\noindent (iii) The field--free DEOM propagation is then carried out
to obtain $\big\{\rho^{(n)}_{\bf n}(t; \hat \mu_{\T})\big\}$,
the DEOM--space correspondence to $e^{-i{\cal L}_{\M}t}(\hat\mu_{\T}\rho^{\rm eq}_{\T})$;

\noindent  (iv) Calculate
\Eq{dipole_corr} in terms of the expectation value like \Eq{PTt_DDOs};
i.e.,
\be\label{dipole_corr_DDOs}
  \la \hat\mu_{\T}(t)\hat\mu_{\T}(0)\ra
={\rm tr}_{\tS}\Big[\hat\mu_{\tS}\rho^{(0)}(t; \hat \mu_{\T})
   +\sum_{ak}\nu^{\B}_a\hat\mu_{\tS}\rho^{(1)}_{ak}(t; \hat \mu_{\T})\Big].
\ee
Note that these formulas are similar but different from that in Ref.\,\onlinecite{Wan1894} due to the different forms of total dipole.

\section{The CODDE formalism}
\label{sec3}

\subsection{Prelude}
The second--order perturbative approach
can be viewed as one--dissipaton theories,
but differ at their resummation treatments on $\{\rho^{(n>1)}_{\bf n}(t)\}$, the higher--order influence.
To be concrete,
the DEOM (\ref{DEOM}) with $n=0$ can be recast as
\be\label{QME0}
 \dot{\rho}_{\tS}(t)=-i{\cal L}(t)\rho_{\tS}(t)
   -i\sum_{ak} \big[\hat Q^{\tS}_{a}-\nu_a\hat \mu_{\tS}\epsilon(t), \rho^{(1)}_{ak}(t)\big].
\ee
 The time--nonlocal formalism, also called
the chronological ordering prescription (COP)
of CS-QDT,  \cite{Muk81509,Yan05187,Xu037,Xu029196}
adopts the simplest one--dissipaton level truncation
by setting all $\rho^{(n>1)}_{\bf n}(t)= 0$.
The resultant DEOM (\ref{DEOM}), with setting $\{\rho^{(n>L)}_{\bf n}=0\}$
at the $L=1$ level, is terminated by
\be\label{dotrho1_cop}
 \dot\rho^{(1)}_{ak}(t)
\approx
  -\big[i{\cal L}(t)+\gamma_k\big]\rho^{(1)}_{ak}(t)
  -i[{\cal C}_{ak}-{\cal D}^\prime_{ak}\epsilon(t)]\rho_{\tS}(t) .
\ee
Its formal solution is
\be\label{rho1_cop}
 \rho^{(1)}_{ak}(t)= \rho^{\COP}_{ak}(t)  +\delta\rho^{\COP}_{ak}(t),
\ee
with
\begin{align} \label{rho_cop}
  \rho^{\COP}_{ak}(t)
&= -i\int^{t}_{-\infty}\!\!
   \d\tau  e^{-\gamma_k(t-\tau)}
  {\cal G}(t,\tau){\cal C}_{ak}\rho_{\tS}(\tau),
\\ \label{delrho_cop}
  \delta\rho^{\COP}_{ak}(t)
&=  i\int^{t}_{-\infty}\!\!\!\d\tau
  e^{-\gamma_k(t-\tau)}{\cal G}(t,\tau){\cal D}^\prime_{ak}\rho_{\tS}(\tau) \epsilon(\tau).
\end{align}
Here, ${\cal G}(t,\tau)$ is the bath--free propagator,
satisfying
\be\label{calG_sys}
 \frac{\partial}{\partial t}{\cal G}(t,\tau)
 =-i{\cal L}(t){\cal G}(t,\tau),
\ \ \text{with ${\cal G}(t,t)=1$}.
\ee

 The time--local CS-QDT is also called
the partial ordering prescription (POP)
formalism.\cite{Muk81509,Yan05187,Xu037,Xu029196}
Within the complete second--order theory, the POP adopts the bath--free approximant of
$\rho_{\tS}(\tau)\approx {\cal G}(\tau,t)\rho_{\tS}(t)$
to \Eq{rho_cop}.
It is noticed that
${\cal G}(t,\tau)\hat O = G(t,\tau)\hat OG^{\dg}(t,\tau)$,
where $G(t,\tau)$ is the Hilbert--space counterpart to
${\cal G}(t,\tau)$.
Consequently,
${\cal G}(t,\tau)\big\{\hat O[{\cal G}(\tau,t)\rho_{\tS}(t)]\big\}
 =[{\cal G}(t,\tau)\hat O]\rho_{\tS}(t)$.
The POP or time--local counterparts to \Eq{rho_cop} is then
\begin{align}\label{rho_POP}
  \rho^{\POP}_{ak}(t)
&= -i\big[\hat X_{ak}(t)\rho_{\tS}(t)
    -\rho_{\tS}(t)\hat X^{\dg}_{a{\bar k}}(t)\big],
\end{align}
where
\begin{align}\label{hatXak}
 \hat X_{ak}(t)
&= \sum_b \eta_{abk} \!  \int^{t}_{-\infty}\!\!
   \d\tau e^{-\gamma_k(t-\tau)}
  {\cal G}(t,\tau)\hat Q^{\tS}_b.
\end{align}
The associated index $\bar k$, the same
as that in \Eq{FF_exp_rev},
is defined via $\gamma_{\bar k}\equiv \gamma^{\ast}_k$.
 By applying the Dyson equation for the
bath--free propagator,
with ${\cal L}(t)\equiv {\cal L}_{\tS}+{\cal L}_\text{sf}(t)$,
\be\label{calG_Dyson}
 {\cal G}(t,\tau)={\cal G}_{\tS}(t-\tau)
  -i\!\int_{\tau}^{t}\!\d\tau'{\cal G}(t,\tau')
  {\cal L}_\text{sf}(\tau'){\cal G}_{\tS}(\tau'-\tau).
\ee
This leads to \Eq{hatXak} the expression,
\be\label{hatXak_sum}
  \hat X_{ak}(t) = \wti X_{ak} +\delta\hat X_{ak}(t),
\ee
with
\be\label{wtiXak}
  \wti X_{ak} = \sum_b \frac{\eta_{abk}}{i{\cal L}_{\tS}+\gamma_k}\hat Q^{\tS}_b.
\ee
The field--driven term, $\delta\hat X_{ak}(t)$,
satisfies
\be\label{dot_del_wtiQ_POP}
 \delta\dot{\hat X}_{ak}(t)
 = -[i{\cal L}(t)+\gamma_k]\delta\hat X_{ak}(t)
  -i{\cal L}_\text{sf}(t)\wti X_{ak}.
\ee
Note that
 ${\cal L}_\text{sf}(t)\hat O\equiv -\epsilon(t)[\hat \mu_{\tS},\hat O]$.

\subsection{CODDE formalism}
The CODDE are based on the following two ansatz. Firstly, we set $\rho^{(1)}_{ak}(t)=\rho^{\POP}_{ak}(t)
  +\delta\rho^{\COP}_{ak}(t)$ [cf.\,\Eqs{rho_POP} and (\ref{delrho_cop})].
The CODDE correspondence to \Eq{rho1_cop},
as inferred from \Eqs{rho_POP}
and (\ref{hatXak_sum}), would read
\be\label{rho1_codde}
 \rho^{(1)}_{ak}(t)
 = -i\big[\wti X_{ak}\rho_{\tS}(t)-\rho_{\tS}(t)\wti X^{\dg}_{a{\bar k}}\big]
  +\varrho_{ak}(t)
\ee
where $\varrho_{ak}(t)$ is defined later; see \Eq{varrho_def}.

  Secondly, we treats $\delta\hat X_{ak}(t)\rho_{\tS}(t)$
with the second--order level linearization ansatz,
\[
 \frac{\d}{\d t}[\delta\hat X_{ak}(t)\rho_{\tS}(t)]
\approx
 \delta\dot{\hat X}_{ak}(t)\rho_{\tS}(t)
-i\delta\hat X_{ak}(t)[{\cal L}(t)\rho_{\tS}(t)].
\]
Together with \Eq{dot_del_wtiQ_POP},
followed by using the commutator identity,
$A({\cal L}B)={\cal L}(AB)-({\cal L}A)B$,
we obtain
\bsube
\begin{align}\label{CODDE_ansatz}
 \frac{\d}{\d t}\big[\delta\hat X_{ak}(t)\rho_{\tS}(t)\big]
&\approx
 -\big[i{\cal L}(t)+\gamma_k\big]\big[\delta\hat X_{ak}(t)\rho_{\tS}(t)\big]
\nl&\quad
   -i\big[{\cal L}_\text{sf}(t)\wti X_{ak}\big]\rho_{\tS}(t),
\\\label{CODDE_ansatzBak}
 \frac{\d}{\d t}\big[\rho_{\tS}(t)\delta\hat X^{\dg}_{a{\bar k}}(t)\big]
&\approx
 -\big[i{\cal L}(t)+\gamma_k\big]
  \big[\rho_{\tS}(t)\delta\hat X^{\dg}_{a{\bar k}}(t)\big]
\nl&\quad
   -i\rho_{\tS}(t)\big[{\cal L}_\text{sf}(t)\wti X^{\dg}_{a{\bar k}}\big].
\end{align}
\esube
Define now
\be\label{varrho_def}
 \varrho_{ak}(t)\equiv -i\big[\delta\hat X_{ak}(t)\rho_{\tS}(t)
  -\rho_{\tS}(t)\delta\hat X^{\dg}_{a{\bar k}}(t)\big]
  +\delta\rho^{\COP}_{ak}(t),
\ee
where 
$\delta\rho^{\COP}_{ak}(t)$ of \Eq{delrho_cop} satisfies
\be\label{dot_delrho_cop}
 \delta{\dot\rho}^{\COP}_{ak}(t)
=-[i{\cal L}(t)+\gamma_k]\delta\rho^{\COP}_{ak}(t)
  +i\epsilon(t){\cal D}'_{ak}\rho_{\tS}(t).
\ee
The equation of motion for $\varrho_{ak}(t)$, \Eq{varrho_def},
can be completed, with  its first term  to be evaluated via the CODDE ansatz,
\Eqs{CODDE_ansatz} and (\ref{CODDE_ansatzBak}).

Finally, the CODDE formalism reads
\bsube\label{CODDE}
\begin{align}
\dot\rho_{\tS} &=\! -\big[i{\cal L}(t)\!+\!{\cal R}_{\tS}(t)\big]\rho_{\tS}
   \!-i\sum_{ak}\!\big[{\cal A}_{a}\!-\!{\cal D}_{a}\epsilon(t)\big]\varrho_{ak},\!
\label{CODDEa} \\
 \dot\varrho_{ak} &= -\big[i{\cal L}(t)+\gamma_k\big]\varrho_{ak}
  -i\epsilon(t)(\wti{\cal C}_{ak}-{\cal D}'_{ak})\rho_{\tS},
\label{CODDEb}
\end{align}
\esube
with
\be\label{calRs}
 {\cal R}_{\tS}(t)\hat O \equiv \sum_{ak}
   \big[\hat Q^{\tS}_{a}-\epsilon(t)\nu_{a}^{\B}\hat\mu_{\tS},\wti X_{ak}\hat O-\hat O\wti X^{\dg}_{a{\bar k}}\,\big],
\ee
and
\be\label{calC_CODDE}
 \wti{\cal C}_{ak}\hat O  \equiv i
  \big[\hat\mu_{\tS},\wti X_{ak}\big]\hat O
      -i \hat O \big[\hat\mu_{\tS},\wti X^{\dg}_{a{\bar k}}\big].
\ee
The above formalism depends locally on the external field $\epsilon(t)$. 
In the absence of external field, 
$\varrho^{\rm eq}_{ak}=\varrho_{ak}(t\rightarrow \pm \infty)=0$ and CODDE is identical to POP formalism, \Eqs{rho_POP}--(\ref{dot_del_wtiQ_POP}).
However, CODDE fixes the
drawback of nonlinearity, \Eq{rho_POP} with \Eq{dot_del_wtiQ_POP}, for the field--dressed dissipation. 
On the other hand, it is well known that COP induces spurious resonances.\cite{Muk81509, Mo05084115, Che09094502}
Overall speaking, CODDE is best among aforementioned three CS-QDTs, with  similar computational cost as COP.

\subsection{Spectroscopies}
 Consider the polarization,
$P_{\T}(t)\equiv {\rm Tr}[\hat\mu_{\T}\rho_{\T}(t)]$,
on the basis of the dissipaton result, \Eq{PTt_DDOs}.
Together with the CODDE's one--dissipaton quantities
of \Eq{rho1_codde}, we obtain
\be\label{PTt_CODDE}
 P_{\T}(t)={\rm tr}_{\tS}\Big[\hat\mu^{\rm eff}_{\tS}\rho_{\tS}(t)
    +\sum_{ak}\nu^{\B}_a\hat \mu_{\tS}\varrho_{ak}(t)\Big],
\ee
where
\be\label{PeffS}
 \hat\mu^{\rm eff}_{\tS} \equiv \hat\mu_{\tS}
\Big[1 - i\sum_{ak} \nu^{\B}_{a}\big(\wti X_{ak}-\wti X^{\dg}_{a{\bar k}}\big)\Big].
\ee
%
 Turn now to the CODDE evaluation on
$\la \hat\mu_{\T}(t)\hat\mu_{\T}(0)\ra$ of \Eq{dipole_corr}.
%
According to \Eq{PTt_CODDE}, we have that
\be\label{muTmuT_CODDE}
 \la \hat\mu_{\T}(t)\hat\mu_{\T}(0)\ra
= {\rm tr}_{\tS}\Big[\hat\mu^{\rm eff}_{\tS}\rho_{\tS}(t;\hat\mu_{\T})
    +\sum_{ak}\nu^{\B}_a \varrho_{ak}(t;\hat\mu_{\T})\Big].
\ee
Here, $\rho_{\tS}(t;\hat\mu_{\T})$ and $\{\varrho_{ak}(t;\hat\mu_{\T})\}$
arise from the field--free CODDE propagation.
The initial conditions are
\be\label{CODDE_correlation_init}
 \begin{bmatrix}
  \rho_{\tS}(0;\hat\mu_{\T}) \\ \varrho_{ak}(0;\hat\mu_{\T})
 \end{bmatrix}
= \begin{bmatrix} \hat \mu_{\tS}+{\cal X}_{\tS}  & \nu_{a}^{\B}\hat\mu_{\tS}
  \\ -\hat W_{ak}  & \hat\mu_{\tS} \end{bmatrix}
  \begin{bmatrix} \rho^{\rm eq}_{\tS} \\  \varrho^{\rm eq}_{ak} \end{bmatrix}.
\ee
where the superoperator ${\cal X}_{\tS}$ denotes the  forward (left) action
component of $i{\cal R}_{\tS}(t) $ in \Eq{CODDE},
\be 
{\cal X}_{\tS}\hat O=-i\nu_{a}^{\B}\hat\mu_{\tS}(\wti X_{ak}\hat O-\hat O\wti X^{\dg}_{a\bar k}).
\ee
This is in line with the left and right action of Hamiltonian.
Similarly, the operator $\hat W_{ak}$ arises from the forward (left) action
component of $(\wti{\cal C}_{ak}-{\cal D}'_{ak})$ in \Eq{CODDE}.
As inferred from \Eqs{calC_CODDE} and (\ref{calD}), we obtain
\be\label{hatW_def}
 \hat W_{ak}= i\big[\hat\mu_{\tS},\wti X_{ak}\big]
  -\sum_{b} \nu^{\B}_b \eta_{ak}\hat \mu_{\tS}.
\ee
This thus completes the CODDE for the non-Condon spectroscopy with the Herzberg--Teller vibronic coupling.

\section{Numerical demonstration}
\label{sec4}
Consider an excitonic dimer system,
\be
  H_{\tS}=\sum_{a=1}^{2}\varepsilon_{a}\hat{B}_{a}^{\dg}\hat{B}_{a}+V(\hat{B}_{1}^{\dg}\hat{B}_{2}+\hat{B}_{2}^{\dg}\hat{B}_{1})+U\hat{B}_{1}^{\dg}\hat{B}_{1}\hat{B}_{2}^{\dg}\hat{B}_{2},
\ee
where $\hat{B}_{a}\equiv |0\ra\la a|$ ($\hat{B}_{a}^{\dag}\equiv |a\ra\la 0|$) are the excitonic annihilation (creation) operators.  $\{\varepsilon_a\}$, $V$, and $U$ are the on-site energies, the interstate coupling, and  the exciton Coulomb interaction. The excitonic system is coupled to the harmonic bath via
\be
  H_{\tS\B}=\sum_{a=1}^{2}\hat{B}_{a}^{\dg}\hat{B}_{a}\hat{F}^{\B}_a.
\ee
The bath spectral density assumes the Brownian oscillator (BO) form
\be\label{Drude}
  J_{11}(\w)=J_{22}(\w)=\frac{2\lambda\zeta\Omega^2\w}{(\w^2-\Omega^2)^2-\zeta^2\w^2},
\ee
and the off-diagonal fluctuations are neglected, i.e., $J_{12}=J_{21}=0$. 
The total system and bath dipole operator takes the form of \Eq{muT0}, where the system dipole operator $\hat{\mu}_{\tS}$ reads
\be
  \hat{\mu}_{\tS}=\mu_{\tS}\sum_{a}(\hat{B}_a^{\dg}+\hat{B}_a),
\ee
and the bath part is $\hat\mu^{\B}_{a}=\mu_{\B}\hat F_{a}^{\B}/(2\lambda)$. Here, $\mu_{\tS}$ and $\mu_{\B}$ are the system and bath dipole strengths, respectively. 
We take $V$ as the reference unit of energy, and set $k_{B}T=\varepsilon_1=\varepsilon_2=U=V$. The BO parameters are given as  $\Omega=V$, $\lambda=0.1V$ and $0.5V$, $\zeta=V$ (underdamped) and $4V$ (overdamped) for comparison.

\begin{figure}
\includegraphics[width=1.0\columnwidth]{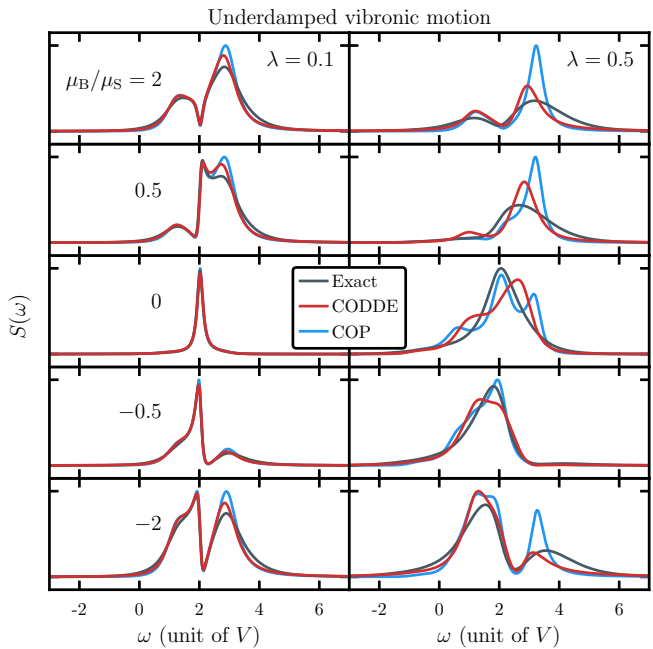}
\caption{Linear absorption spectra $S(\w)$ calculated by \Eq{SA} in the underdamped BO case ($\zeta=V$) for various values of $\mu_{\B}/\mu_{\tS}$. Left panel: $\lambda=0.1V$; Right panel: $\lambda=0.5V$.
}\label{fig1}
\end{figure}

The linear absorption spectra of the total matter are obtained via
\be\label{SA}
  S(\w)={\rm Re}\!\int_{0}^{\infty}\!\!\d t\,e^{i\w t}\la \hat\mu_{\T}(t)\hat\mu_{\T}(0)\ra.
\ee
In \Fig{fig1}, we present the linear absorption spectra in the underdamped BO case ($\zeta=V$). As the bath optical activity is switched on and tuned up, $\mu_{\B}/\mu_{\tS}$ varies from $-2$ to $2$.
The results of the weak coupling case ($\lambda=0.1 V$) are exhibited in the left panel, where both COP and CODDE results are close to exact ones. The exact results are obtained from the converged DEOM calculations. In the right panel, we adopt $\lambda=0.5V$ for the strong vibronic coupling. COP qualitatively deviates from the exact ones, exhibiting spurious peaks.
It is found that CODDE always behaves much better than COP.

\begin{figure}
\includegraphics[width=1.0\columnwidth]{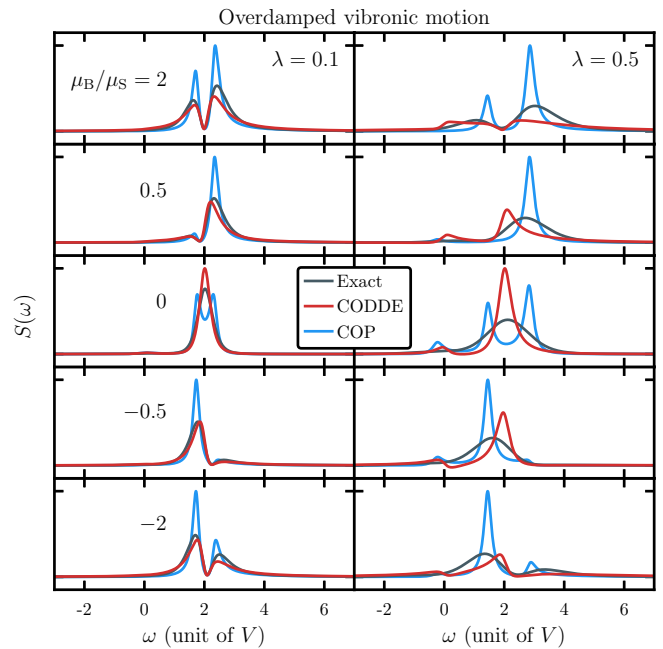}
\caption{Linear absorption spectra $S(\w)$ calculated by \Eq{SA} in the overdamped BO case ($\zeta=4V$) for various values of $\mu_{\B}/\mu_{\tS}$. Left panel: $\lambda=0.1V$; Right panel: $\lambda=0.5V$.  
}\label{fig2}
\end{figure}

Moreover, \Fig{fig2}  exhibits the linear absorption spectra in the overdamped BO case ($\zeta=4V$). 
We observe that even in the weak coupling case (left panel), COP is ill--behaved. CODDE would be the choice of CS-QDT,  although it quantitatively deviate from the exact ones obtained via DEOM when the coupling is strong.

\section{Summary}
\label{sec5}
In summary, we extend the CODDE, an optimized CS-QDT, to deal with the Herzberg--Teller vibronic couplings in dipole--field interactions, with the aid of DEOM formalism.
As CODDE is formally an one--dissipaton theory,  we can extract not only the reduced system dynamics but also the hybridized bath informations.
%
%
The present theory can be readily extended to
the nonequilibrium setups with multiple coupling environment baths.
The heat transport current and noise spectra can also be evaluated.

\begin{acknowledgments}
Support from 
 the Ministry of Science and Technology of China (Nos.\ 2017YFA0204904 and 2021YFA1200103)
 and 
the National Natural Science Foundation of China (Nos.\ 22103073 and 22173088)  
 is gratefully acknowledged. Y. Wang  and Z. H. Chen thank also the
partial support from GHfund B (20210702).
\end{acknowledgments}


\end{document}